\documentstyle[aps]{revtex}
\begin{document}
\title{A New Proof of The Strong Subadditivity Theorem}
\author{Yong-Jian Han, Yong-Sheng Zhang\thanks{%
Electronic address: yshzhang@ustc.edu.cn}, Guang-Can Guo\thanks{%
Electronic address: gcguo@ustc.edu.cn}}
\address{Key Laboratory of Quantum Information, University of Science and Technology\\
of China, CAS, Hefei 230026, People's Republic of China\bigskip \bigskip }
\maketitle

\begin{abstract}
\baselineskip12pt It is well known that the strong subadditivity theorem is
hold for classical system, but it is very difficult to prove that it is hold
for quantum system. The first proof of this theorem is due to Lieb by using
the Lieb's theorem. Here we use the conditions obtained in our previous work
of matrix analysis method{\it \ }to give a new proof of this famous theorem.
This new proof is very elementary, it only needs to carefully analyse the
minimal value of a function. This proof also shows that the conditions
obtained in our previous work{\it \ }are stronger than the strong
subadditivity theorem.

PACS number(s): 03.67.-a, 02.10.Yn, 89.70.+c\medskip 
\end{abstract}

\baselineskip12pt

\section{Introduction}

Entropy is an important concept not only for physics but also for
information science. From the definition of entropy, we can get some simple
properties of it, such as concavity, continuity property, additivity and
subadditivity\cite{wehrl}. But some other properties is not so obvious, such
as the strong subadditivity theorem (SSA). Among all of the properties of
entropy, the most famous one is the SSA, and it is very difficult to prove
this theorem for quantum system. The content of this theorem can be
expressed as the following: two overlapping subsystem $AB$ and $BC$, the
entropy of their union ($ABC$) plus the entropy of their intersection ($B$)
does not exceed the sum of the entropies of the subsystems ($AB$ and $BC$)%
\cite{preskill}, that is 
\begin{equation}
S(\rho _{ABC})+S(\rho _B)\leq S(\rho _{AB})+S(\rho _{BC}).  \eqnum{1}
\end{equation}
where $S(\rho )=-Tr(\rho ln\rho ).$ It is well known that this theorem is
true for classical information theory, but to prove this theorem is true for
quantum system is very difficult. This theorem is first conjectured to be
true for the quantum system by Lanford and Robibson\cite{LR}. The first
proof of this conjecture is given by Lieb {\it et al. }several years later.
This proof is based on the concave of the function $S(\rho _{12})-S(\rho _1)$
in $\rho _{12}$\cite{lieb}. Another proof based on the same fact was
proposed by Uhlmann\cite{ulhmann,ulhmann1}.

Recently, quantum information theory attracts more and more attentions for
its misterious properties and its potential applications in science and
technology\cite{tele,ekert}. The SSA plays an important role in this new
field\cite{neilson} too. The fundamental scource in quantum information is
entanglement between many particles which can be viewed as the relations
between the partial particles. So distinguish whether a set of the partial
particles come from a single state ($N$-representability problem)\cite
{erhal,colemn} and further to obtain its entanglement property are important
in quantum information while SSA gives a strong constraint on the partial
particles and the whole system. The convenience of the SSA is that it has
explicit physical and manipulating meaning. So it is a convenient necessary
criterion for the $N$-representability problem. Recently, we use the matrix
analysis method\cite{han} to get some necessary conditions for the $N$%
-presentability problem. We find that using these conditions we can get a
new proof for the SSA. Our new proof is elementary, we need only to use the
Lagrange multiplier method and carefully analyse the minimum of a function.

\section{The Theorem and the Proof}

There is a density matrix $\rho _{ABC\text{,}}$ where the particle $A,B$ and 
$C$ are in $L-$dimension, $M-$dimension and $N-$dimension Hilbert space,
respectively. Let $\{\lambda _{AB}^{(1)},\lambda _{AB}^{(2)},\cdots ,\lambda
_{AB}^{(LM)}\},$ $\{\lambda _{BC}^{(1)},\lambda _{BC}^{(2)},\cdots ,\lambda
_{BC}^{(MN)}\},$ $\{\lambda _B^{(1)},\lambda _B^{(2)},\cdots ,\lambda
_B^{(M)}\}$ and $\{\lambda _{ABC}^{(1)},\lambda _{ABC}^{(2)},\cdots ,\lambda
_{ABC}^{(LMN)}\}$ are the eigenvalues of the density matrix $\rho _{AB},\rho
_{BC},\rho _B$ and $\rho _{ABC}$, respectively (where $\rho _{AB},\rho _{BC}$
and $\rho _B$ are gotten by tracing the other particles from $\rho _{ABC}$),
and they are arranged in increasing order. We defined vectors $\lambda
_{AB}=\{\lambda _{AB}^{(1)},$ $\lambda _{AB}^{(2)},$ $\cdots ,$ $\lambda
_{AB}^{(LM)}\},$ $\lambda _{BC}=\{\lambda _{BC}^{(1)},$ $\lambda
_{BC}^{(2)}, $ $\cdots ,$ $\lambda _{BC}^{(MN)}\},$ $\lambda _B=\{\lambda
_B^{(1)},$ $\lambda _B^{(2)},$ $\cdots ,$ $\lambda _B^{(M)}\},$ $\lambda
_{ABC}=\{\lambda _{ABC}^{(1)},$ $\lambda _{ABC}^{(2)},$ $\cdots ,$ $\lambda
_{ABC}^{(LMN)}\}$ and $\lambda _{AB}^B=\{\sum_{i=1}^L\lambda _{AB}^{(i)},$ $%
\sum_{i=L+1}^{2L}\lambda _{AB}^{(i)},$ $\cdots ,$ $\sum_{i=L(M-1)+1}^{LM}%
\lambda _{AB}^{(i)}\},$ $\lambda _{BC}^B=\{\sum_{i=1}^N\lambda _{BC}^{(i)},$ 
$\sum_{i=N+1}^{2N}\lambda _{BC}^{(i)},$ $\cdots ,$ $\sum_{i=(M-1)N+1}^{MN}%
\lambda _{AB}^{(i)}\},$ $\lambda _{ABC}^{AB}=\{\sum_{j=1}^N\lambda
_{ABC}^{(j)},$ $\sum_{j=N+1}^{2N}\lambda _{ABC}^{(j)},$ $\cdots ,$ $%
\sum_{j=(LM-1)N+1}^{LMN}\lambda _{ABC}^{(j)}$ $\},$ and $\lambda
_{ABC}^{BC}= $\{$\sum_{j=1}^L\lambda _{ABC}^{(j)},$ $\sum_{j=L+1}^{2L}%
\lambda _{ABC}^{(j)},$ $\cdots ,$ $\sum_{j=(MN-1)L+1}^{LMN}\lambda
_{ABC}^{(j)}$ $\}. $ Using the matrix analysis method, we get the following
two lemmas on the eigenvalues\cite{han}.

{\bf Lemma 1}. Using the notes defined before, we can get the relations
between the eigenvalues of $\rho _{BC}$, $\rho _{AB\text{,}}$ $\rho _B$ and $%
\rho _{ABC}$ as 
\begin{equation}
\lambda _{ABC}^{AB}\succ \lambda _{AB}  \eqnum{2.1}
\end{equation}
\begin{equation}
\lambda _{ABC}^{BC}\succ \lambda _{BC}  \eqnum{2.2}
\end{equation}
\begin{eqnarray}
\lambda _{BC}^B &\succ &\lambda _B  \eqnum{2.3.1} \\
\lambda _{AB}^B &\succ &\lambda _B  \eqnum{2.3.2}
\end{eqnarray}

{\bf Lemma 2}. Suppose $rank(\rho _{ABC})=LMN-Ls,$ $rank(\rho _{BC})=MN-s,$ $%
rank(\rho _{AB})=LM-r$ and $rank(\rho _B)=M-t,$ if $r$ and $s$ satisfy the
condition $Nr\leq Ls,$ there will be 
\begin{equation}
t\leq [\frac{r-1}L]+1,  \eqnum{3}
\end{equation}
where $[x]$ is the maximum integer which is smaller than $x$.

The notation $y\succ x$ mean that the vector $x$ is majorized by the vector $%
y$. The majorization is defined as the following. Let $x=\{x_1,x_2,\cdots
,x_n\}$ and $y=\{y_1,y_2,\cdots ,y_n\}$ are $n$-dimensional vectors and the
elements are arranged in increasing order. Then the vector $x$ is majorized
by vector $y$\cite{olkin}, denoted by $y\succ x,$ if for each $k$ $%
(k=1,2,\cdots n)$ the following inequality is hold 
\[
\sum_{i=1}^kx_i\geq \sum_{i=1}^ky_i 
\]
and the equality is hold when $k=n.$ Under these two Lemmas, we can find
that the SSA is hold in the following.

{\bf Theorem } There are four normalized vectors $\lambda ^{ABC}=\{\lambda
_1^{ABC},\lambda _2^{ABC},\cdots ,\lambda _{LMN}^{ABC}\},$ $\lambda
^{AB}=\{\lambda _1^{AB},\lambda _2^{AB},\cdots ,\lambda _{LM}^{AB}\},$ $%
\lambda ^{BC}=\{\lambda _1^{BC},\lambda _2^{BC},\cdots ,\lambda _{MN}^{BC}\}$
and $\lambda ^B=\{\lambda _1^B,\lambda _2^B,\cdots ,\lambda _M^B\}$, the
elements of these vectors are non-negative and arranged in increasing and
define the vectors $\lambda _{AB}^{ABC},$ $\lambda _{BC}^{ABC}$ and $\lambda
_B^{BC},$ $\lambda _B^{AB},$ which are similar to the vectors in lemma 1. If
the elements of these vectors satisfy the following conditions

1. $\lambda _{AB}^{ABC}\succ \lambda ^{AB}$ $;$

2. $\lambda _{BC}^{ABC}\succ \lambda ^{BC};$

3. $\lambda _B^{BC}\succ \lambda ^B$ and $\lambda _B^{AB}\succ \lambda ^B;$

4. Suppose the vector $\lambda =\lambda ^{ABC}$ has only $Ls$ zero elements
and $\lambda ^{BC}$ has $s$ zero elements, and if the vector $\lambda ^{AB}$
has $r$ zero elements, there are at least $[\frac{r-1}L]+1$ elements of the
vector $\lambda ^B$ are zeroes. If exchange the role of the vector $\lambda
^{AB}$ and $\lambda ^{BC},$ the similar result must be hold also.

Thus the following inequality is hold 
\begin{equation}
S(\lambda )+S(\lambda ^B)\leq S(\lambda ^{BC})+S(\lambda ^{AB}),  \eqnum{4}
\end{equation}
where $S(\lambda )=\sum_{i=1}^{LMN}(-\lambda _i\ln \lambda _i).$

The proof of the theorem is technical. We use the Lagrange multiplier method
to get the minimal value of a function under the conditions 1, 2 and 3.
Because there are many possible extreme points, we need to find out the
minimal one. We use some facts to find that when the function gets the
minimal value, all of the nonzero elements are equal to each other. Then use
the condition 4 to get the minimal value of the function.

{\it Proof. }At first, we define a function 
\begin{equation}
F=\sum_{i=1}^{LM}(-\lambda _i^{AB}\ln \lambda
_i^{AB})+\sum_{i=1}^{MN}(-\lambda _i^{BC}\ln \lambda
_i^{BC})+\sum_{i=1}^M(\lambda _i^B\ln \lambda _i^B)+\sum_{i=1}^{LMN}(\lambda
_i\ln \lambda _i),  \eqnum{5}
\end{equation}
If we can prove that the minimum of this function is not less than $0$ under
the conditions 1-4, the theorem is true. So the proof becomes to find the
minimal value of a function under some conditions. Obviously, the minimal
value of this function exists and is finite. Now we use the Lagrange
multiplier method to deal with the conditions 1, 2, 3 and define a new
function

\begin{eqnarray}
G &=&\sum_{i=1}^{LM}(-\lambda _i^{AB}\ln \lambda
_i^{AB})+\sum_{i=1}^{MN}(-\lambda _i^{BC}\ln \lambda
_i^{BC})+\sum_{i=1}^M(\lambda _i^B\ln \lambda _i^B)+\sum_{i=1}^{LMN}(\lambda
_i\ln \lambda _i)  \eqnum{6} \\
&&+\sum_{i=1}^{LM-1}\alpha _i^1(\sum_{j=1}^i\lambda
_j^{AB}-\sum_{j=1}^{Ni}\lambda _j-x_{1i}^2)+\sum_{k=1}^{MN-1}\beta
_k^1(\sum_{j=1}^k\lambda _j^{BC}-\sum_{j=1}^{Lk}\lambda _j-y_{1k}^2) 
\nonumber \\
&&+\sum_{i=1}^{M-1}\alpha _i^2(\sum_{j=1}^{Li}\lambda
_j^B-\sum_{j=1}^i\lambda _j^{AB}-x_{2i}^2)+\sum_{k=1}^{M-1}\beta
_k^2(\sum_{j=1}^{Nk}\lambda _j^B-\sum_{j=1}^k\lambda _j^{BC}-y_{2k}^2) 
\nonumber \\
&&+\sum_{i=0}^{LMN}\gamma _i(\lambda _{i+1}-\lambda
_i-z_i^2)+\sum_{i=1}^{LM}u_i(\lambda _{i+1}^{AB}-\lambda _i^{AB}-r_i^2) 
\nonumber \\
&&+\sum_{i=1}^{MN}v_i(\lambda _{i+1}^{BC}-\lambda
_i^{BC}-s_i^2)+\sum_{i=1}^Mw_i(\lambda _{i+1}^B-\lambda _i^B-t_i^2) 
\nonumber \\
&&+a_1(\sum_{i=1}^{LM}\lambda _i^{AB}-1)+a_2(\sum_{i=1}^{MN}\lambda
_i^{BC}-1)+a_3(\sum_{i=1}^M\lambda _i^B-1)+a_4(\sum_{i=1}^{LMN}\lambda _i-1),
\nonumber
\end{eqnarray}
where the parameters $\alpha _i^j,$ $\beta _k^j,$ $u_i,$ $v_i$ and $w_i,$ $%
\gamma _j,$ $a_i$ are Lagrange multipliers, $x_{1i}^2,$ $y_{1k}^2,$ $%
x_{2i}^2,$ $y_{2k}^2,$ $z_i^2,$ $r_i^2,$ $s_i^2$ and $t_i^2$ are introduced
to make the inequalities to be equations. We have used the conditions that $%
\lambda _i^{AB},\lambda _j^B,\lambda _k^{BC}$ and $\lambda _i$ are arranged
in increasing order and let $\lambda _0=0$.

Then when $G$ get the minimal value, there must be some constraints on the
parameters and variables. First, we can get $\alpha _i^jx_{ji}=0$ $%
(i=1,2,\cdots ,LM-1;$ $j=1,2)$ and the similar relations between $\beta _k^j$
and $y_{jk}$, $\gamma _i$ and $z_i,$ $u_i$ and $r_i,v_i$ and $s_i$, $w_i$
and $t_i.$ The second, we get the relations between the elements of the
vector $\lambda ^{AB}$ and $\lambda ,$%
\begin{equation}
\sum_{j=1}^i\lambda _j^{AB}-\sum_{j=1}^{Ni}\lambda _j=x_{1i}^2,i=1,2,\cdots
,LM-1  \eqnum{7}
\end{equation}
and the similar relations between $\lambda $ and $\lambda ^{BC},$ $\lambda
^{AB}$ and $\lambda ^B,$ $\lambda ^{BC}$ and $\lambda ^B.$ Then the
relations between the vector $\lambda ^{AB}$ can be gotten 
\begin{eqnarray}
\lambda _{i+1}^{AB}-\lambda _i^{AB} &=&r_i^2,i=1,\cdots ,LM-1;  \eqnum{8} \\
\sum_{i=1}^{LM}\lambda _i^{AB}-1 &=&0,  \nonumber
\end{eqnarray}
and the similar relations between the vectors of $\lambda ^{BC},\lambda ^B$
and $\lambda .$

The most important constraints are the equations between the vectors $%
\lambda _i^{AB},\lambda _i^{BC}$ $,\lambda _i^B$and $\lambda _i$%
\begin{equation}
-\ln \lambda _i^{AB}-1+\sum_{j=i}^{LM-1}\alpha _i^1-\sum_{j=[\frac{i-1}L%
]+1}^{M-1}\alpha _j^2+u_{i-1}-u_i+a_1=0\text{ }(i=1,2,\cdots ,LM). 
\eqnum{9.1}
\end{equation}
\begin{equation}
-\ln \lambda _i^{BC}-1+\sum_{j=i}^{MN-1}\beta _j^1-\sum_{j=[\frac{i-1}N%
]+1}^{M-1}\beta _j^2+v_{i-1}-v_i+a_2=0\text{ }(i=1,2,\cdots ,MN). 
\eqnum{9.2}
\end{equation}
\begin{equation}
\ln \lambda _i^B+1+\sum_{j=i}^{M-1}\alpha _i^2+\sum_{j=i}^{M-1}\beta
_i^2+w_{i-1}-w_i+a_3=0\text{ }(i=1,2,\cdots ,M).  \eqnum{9.3}
\end{equation}
\begin{equation}
\ln \lambda _i+1-\sum_{j=[\frac{i-1}N]+1}^{LM-1}\alpha _j^1-\sum_{j=[\frac{%
i-1}L]+1}^{MN-1}\beta _j^1+\gamma _{i-1}-\gamma _i+a_4=0\text{ }%
(i=1,2,\cdots ,LMN).  \eqnum{9.4}
\end{equation}

Since the number of the possible cases are so large, it is very difficult to
get the solutions directly. We point out some useful facts to reduce the
possible solutions and to find the minimal value of the function $G$.

{\bf Fact 1. }When the function $G$ get the minimum, suppose that parameters 
$\alpha _i^1$ and $\alpha _j^2$ are the nearest nonzero parameter act on the
elements of vector $\lambda ^{AB}$, if the parameters $u_i$ and $u_{Lj}$ are
zeroes, all the parameter $u_k$ $(i\leq k\leq Lj)$ are equal to zeroes. This
fact is true for the other parameters $v_i,w_i,\gamma _i.$

{\it Proof}. Without loss of generality, we only consider the parameter $%
u_i. $ Suppose the fact is not true, there are some parameters $u_p$ $%
(i<m\leq p\leq n<Lj)$ are not zeroes. For simplicity, we suppose there are
no more nonzero parameters $\alpha _i^1$ and $\alpha _j^2$. Then we get the
conditions from (9.1) 
\begin{eqnarray}
-\ln \lambda _m^{AB}-1+\alpha _i^1-\alpha _j^2+0-u_m+a_1 &=&0,  \nonumber \\
-\ln \lambda _{m+1}^{AB}-1+\alpha _i^1-\alpha _j^2+u_m-u_{m+1}+a_1 &=&0, 
\nonumber \\
&&\vdots  \eqnum{10} \\
-\ln \lambda _n^{AB}-1+\alpha _i^1-\alpha _j^2+u_{n-1}-u_n+a_1 &=&0, 
\nonumber \\
-\ln \lambda _{n+1}^{AB}-1+\alpha _i^1-\alpha _j^2+u_n-0+a_1 &=&0,  \nonumber
\end{eqnarray}
where we have used the conditions that the parameters $u_{m-1}$ and $u_{n+1%
\text{ }}$are zeros. Since the parameters $u_p$ $(i<m\leq p\leq n<Lj)$ are
nonzero, then we get $\lambda _m^{AB}=\lambda _{m+1}^{AB}=\cdots =\lambda
_n^{AB}=\lambda _{n+1}^{AB}.$ So we have the relations between these nonzero
parameters 
\begin{equation}
-u_m=u_m-u_{m+1}=\cdots =u_{n-1}-u_n=u_n,  \eqnum{11}
\end{equation}
that is, $u_n=(n-m+1)u_m=-u_m.$ So $u_m=0$, then all of the parameters $u_p$ 
$(i<m\leq p\leq n<Lj)$ are zeroes, which is inconsistent with our suppose.
So this fact is true. QED.

Since the fact 1, the parameters $u_k$ affect the result only when there are
some nonzero parameter $\alpha _i^1$ or $\alpha _j^2$ make $k=i$ or $k=Lj$.
For this situation, we have the following fact.

{\bf Fact 2}. When the function $G$ get the minimum, if there are a set of
parameters $u_k$ $(m\leq k\leq n)$ are nonzero and there are some nonzero
parameters $\alpha _i^1$ make $m\leq i\leq n$, This situation is equal to
the situation where the parameters $u_k$ $(m\leq k\leq n)$ and $\alpha _i^1$
are zeroes, but two new parameters $\alpha _{m-1}^1$ and $\alpha _{n+1}^1$
should be added, and the parameters $\gamma _i$ $(0\leq i\leq LMN)$ should
be adjusted.

{\it Proof}. Without loss of generality, we suppose only the nonzero
parameter $\alpha _i^1$ satisfy the condition $m\leq i\leq n.$ For
simplicity, we suppose there is no other nonzero parameters act on the
eigenvalues of $\rho _{AB}.$ Since the parameter $\alpha _i^1$ are nonzero,
then 
\begin{equation}
\sum_{j=1}^i\lambda _j^{AB}-\sum_{j=1}^{Ni}\lambda _j=0;  \eqnum{12}
\end{equation}

and the parameters $u_k$ $(m\leq k\leq n)$ are nonzero, we get 
\begin{equation}
\lambda _m^{AB}=\lambda _{m+1}^{AB}=\cdots =\lambda _n^{AB}=\lambda
_{n+1}^{AB}.  \eqnum{13}
\end{equation}

Since we have the condition $\sum_{j=1}^{i-1}\lambda _j^{AB}\geq
\sum_{j=1}^{N(i-1)}\lambda _j$, together with the equation (12), we get $%
\lambda _i^{AB}\leq \sum_{j=N(i-1)+1}^{Ni}\lambda _j.$ On the other hand, $%
\lambda _{i+1}^{AB}\geq \sum_{j=Ni+1}^{N(i+1)}\lambda _j$, that is, $%
\sum_{j=N(i-1)+1}^{Ni}\lambda _j\geq \sum_{j=Ni+1}^{N(i+1)}\lambda _j.$
Because of the condition $\lambda _{N(i-1)+1}\leq \lambda _{N(i-1)+2}\leq
\cdots \leq \lambda _{N(i+1)},$ we get the equation $\lambda
_{N(i-1)+1}=\lambda _{N(i-1)+1}=\cdots =\lambda _{N(i+1)}.$ So we get $%
\lambda _i^{AB}=\sum_{j=N(i-1)+1}^{Ni}\lambda _j$ and $\lambda
_{i+1}^{AB}=\sum_{j=Ni+1}^{N(i+1)}\lambda _j,$ that is 
\begin{equation}
\sum_{j=1}^{i-1}\lambda _j^{AB}-\sum_{j=1}^{N(i-1)}\lambda
_j=0,\sum_{j=1}^{i+1}\lambda _j^{AB}-\sum_{j=1}^{N(i+1)}\lambda _j=0. 
\eqnum{14}
\end{equation}

Continue to use this method we can get 
\begin{equation}
\sum_{j=1}^{m-1}\lambda _j^{AB}-\sum_{j=1}^{N(m-1)}\lambda
_j=0,\sum_{j=1}^{n+1}\lambda _j^{AB}-\sum_{j=1}^{N(n+1)}\lambda _j=0. 
\eqnum{15}
\end{equation}

From these equations, we can find this is just as there are two nonzero
parameters $\alpha _{m-1}^1$ and $\alpha _{n+1}^1,$ and the nonzero
parameter $\alpha _i^1$ have no effect in this case. From the constraints on 
$\lambda _k^{AB}$, there wll be 
\begin{eqnarray}
-\ln \lambda _k^{AB}-1+\alpha _i^1+a_1 &=&0\text{ }\left( 1\leq k\leq
m-1\right) ,  \eqnum{16.1} \\
-\ln \lambda _k^{AB}-1+\alpha _i^1+u_{k-1}-u_k+a_1 &=&0\text{ }\left( m\leq
k\leq i\right) ,  \eqnum{16.2} \\
-\ln \lambda _k^{AB}-1+u_{k-1}-u_k+a_1 &=&0\text{ }\left( i+1\leq k\leq
n+1\right) ,  \eqnum{16.3} \\
-\ln \lambda _k^{AB}-1+a_1 &=&0\text{ }\left( n+2\leq k\leq LM\right) . 
\eqnum{16.4}
\end{eqnarray}

Since the elements $\lambda _k^{AB}$ $(m\leq k\leq n+1)$ are equal to each
other, then $u_m=u_m-u_{m+1}=\cdots =u_{i-1}-u_i\equiv \alpha $, $%
u_i-u_{i+1}=u_{i+1}-u_{i+2}=\cdots =u_n\equiv \beta $ and $\beta =\alpha
+\alpha _i^1$. If we let the nonzero parameters $\alpha _{m-1}^1=-\alpha $
and $\alpha _{n+1}^1=\beta $, the equations are the same. Now we consider
the effect of this substitution on the vector $\lambda .$ For simplicity, we
suppose also that there are only the nonzero parameter $\alpha _i^1$ act on
the vector $\lambda .$ Then the equations are 
\begin{eqnarray}
\ln \lambda _k+1-\alpha _i^1+\gamma _{k-1}-\gamma _k+a_4 &=&0\text{ }\left(
1\leq k\leq Ni\right) ,  \eqnum{17.1} \\
\ln \lambda _k+1+\gamma _{k-1}-\gamma _k+a_4 &=&0\text{ }\left( Ni+1\leq
k\leq LMN\right) .  \eqnum{17.2}
\end{eqnarray}

Insert the parameters $\alpha _{m-1}^1$ and $\alpha _{n+1}^1$ into the
equations, we can find that 
\begin{eqnarray}
\ln \lambda _k+1-\alpha _{m-1}^1-\alpha _{n+1}^1+\gamma _{k-1}-\gamma _k+a_4
&=&0\text{ }\left( 1\leq k\leq Ni\right) ,  \eqnum{18.1} \\
\ln \lambda _k+1-\alpha _{n+1}^1+\gamma _{k-1}^{^{\prime }}-\gamma
_k^{^{\prime }}+a_4 &=&0\text{ }\left( Ni+1\leq k\leq N(n+1)\right) 
\eqnum{18.2} \\
\ln \lambda _k+1+\gamma _{k-1}^{^{\prime }}-\gamma _k^{^{\prime }}+a_4 &=&0%
\text{ }\left( N(n+1)+1\leq k\leq LMN\right)  \eqnum{18.3}
\end{eqnarray}
where $\gamma _k^{^{\prime }}$ $(Ni+1\leq k\leq LMN)$ are the new parameters
to make the equations are the same as the equation (17). This is just as the
situation that the parameters $\alpha _{m-1}^1$ and $\alpha _{n+1}^1$ are
nonzero, and the parameter $\gamma _k$ is adjusted. QED

This fact is also true for the parameters $\beta _j^1.$ This fact tell us
that any solution found in the former situation can be found in the later
case. In the following, we always suppose we have already done this change.
After making these changes there is no nonzero parameters $\alpha _i^1$ or $%
\beta _j^1$ make the parameter $u_i(v_j)$ nonzero and the parameters $\gamma
_i$ are substituted by $\gamma _i^{^{\prime }}.$

{\bf Fact 3}. When the function $G$ get the minium, if $i$ and $j$ are the
nearest indexes to make the equations $\sum_{k=1}^i\lambda
_k^{AB(BC)}=\sum_{k=1}^{N(L)i}\lambda _k$ and $\sum_{k=1}^j\lambda
_k^{BC(AB)}=\sum_{k=1}^{L(N)j}\lambda _k$ to be hold, the elements of the
vector $\lambda $ between $Ni$ and $Lj$ are equal to each other.

{\it Proof}. We suppose this conclusion is not true, without loss of
generality, let $Lj>Ni$. Then there are some elements satisfy the following
conditions $\lambda _{Lj}=\lambda _{Lj-1}=\cdots =\lambda _p\equiv \lambda
_b>\lambda _a\equiv \lambda _{Ni}=\lambda _{Ni+1}=\cdots =\lambda _q$, for
simplification, we suppose that $Lj-p\geq q-Ni.$ If we define the following
parameters $\Delta _l$ and $\Delta _m^{^{\prime }}$ as $\sum_{k=i+1}^l%
\lambda _k^{AB}-\sum_{k=Ni+1}^{Nl}\lambda _k=\Delta _l$ ($i+1\leq l\leq
\left[ \frac{Lj}N\right] ,[x]$ is the maximal integer which is smaller than $%
x$) and $\sum_{k=Nm+1}^{Nj}\lambda _k-\sum_{k=m}^j\lambda _k^{BC}=\Delta
_m^{^{\prime }}$ ($\left[ \frac{Ni}L\right] \leq m\leq j$), we can find that
all of these parameters are more than zero. Then we take out the minimal
number from $\frac{\Delta _l}{l-i}$ and $\frac{\Delta _m^{^{\prime }}}{j-m}$%
, we denote it by $\Delta $, obviously it is more than zero. Now we change
the element $\lambda _2$ by $\lambda _2-\frac{\Delta ^{^{\prime }}}{Lj-p}$
and $\lambda _1$ by $\lambda _1+\frac{\Delta ^{^{\prime }}}{q-Ni}$ where the
parameter $\Delta ^{^{\prime }}=(q-Ni)\Delta .$ After these substitution,
the new elements of the vector $\lambda ^{^{\prime }}$ satisfy all of the
conditions. The entropy of the vector $\lambda ^{^{\prime }}$ is larger than
the entropy of the vector $\lambda $ and the entropy of the other vector is
invariable. So the function $G$ for the new vector is smaller than the
former which is inconsistent with the suppose. QED.

This fact is also true for the vector $\lambda ^B.$ Since we have the fact
3, then we want to know how many nonzero parameters $\alpha _i^1$ and $\beta
_j^1$ in the section where all of the elements are the same. We have the
following fact

{\bf Fact 4}. When the function $G$ get the minimum, there is no nonzero
parameters $\alpha _i^1$ and $\beta _j^1$ in the section where all of the
elements of vector $\lambda $ are the same except for the edge parameters.

{\it Proof}{\bf . }We first point out that there are at most four nonzero
parameters $\alpha _i^1$ or $\beta _j^1$ in the section where all of the
elements of vector $\lambda $ are equal to each other if the conclusion is
not true. If this assert is not true, there are at least five nonzero
parameters act on the section where all of the elements of the vector $%
\lambda $ are the same. So at least three of them (such as $\alpha _l^1$ $%
(l=i,j,k\cdots )$ or $\beta _l^1$ $(l=i,j,k\cdots )$) are act on the same
vector. Without loss of generality, we suppose there are three nonzero
parameters $\alpha _l^1$ $(l=i,j,k).$ Since the elements $\lambda
_{Ni+1}=\lambda _{Ni+2}=\cdots =\lambda _{Nj}=\lambda _{Nj+1}=\lambda
_{Nj+2}=\cdots =\lambda _{Nk}\equiv \lambda _a.$ We have $\lambda
_{i+1}^{AB}\geq N\lambda _a$ and $\lambda _k^{AB}\leq N\lambda _a,$ since $%
\lambda _{i+1}^{AB}\leq \lambda _k^{AB},$ then all of the elements $\lambda
_l^{AB}$ $(i+1\leq l\leq k)$ are equal to each other. Because we have
already done the changes in the fact 2, and use the fact 1, we find all of
the parameters $u_k$ $(i+1\leq l\leq k)$ are zeroes. Further more, the
parameters $\alpha _j^1$ are zeroes too. So the number of the nonzero
parameters is no more than four, and they divide the section where all the
elements are equal into three smaller sections.

Now we only need to prove the case that less than five parameters are also
zeroes. If these parameters are nonzero and set on the vectors as figure 1,
which makes the function $G$ get the minimum, we take some sufficient small
value $\Delta $ from the elements of the first section to the third section.
At the same time, $\Delta $ must be taken from the left side section of the
parameters $k$ and $j$ to the right side section. Using the same method of
the proof of the fact 3, if the $\Delta $ is sufficient small, all the
conditions will be satisfied. From the following calculating, we can find
that through this manipulation the function $G$ is smaller which is
inconsistent with the minimal suppose.

Let the elements of the vectors before the manipulating are $\lambda
_{Ni+1}=\lambda _{Ni+2}=\cdots =\lambda _{Lk}\equiv \lambda _a,\lambda
_{Nj+1}=\lambda _{Nj+2}=\cdots =\lambda _{Ll}\equiv \lambda _b;\lambda
_j^{AB}=\lambda _{j-1}^{AB}=\cdots =\lambda _s^{AB}\equiv \lambda _a^{AB},$ $%
\lambda _{j+1}^{AB}=\lambda _{j+2}^{AB}=\cdots =\lambda _t^{AB}\equiv
\lambda _b^{AB};\lambda _k^{BC}=\lambda _{k-1}^{BC}=\cdots =\lambda
_u^{BC}\equiv \lambda _a^{BC},\lambda _{k+1}^{BC}=\lambda _{k+2}^{BC}=\cdots
=\lambda _v^{BC}\equiv \lambda _b^{BC}.$ After the manipulate, the new
elements are $\lambda _a^{^{\prime }}=\lambda _a-\frac \Delta {Lk-Ni}%
,\lambda _b^{^{\prime }}=\lambda _a+\frac \Delta {Ll-Nj};$ $\lambda
_a^{AB^{\prime }}=\lambda _a^{AB}-\frac \Delta {j-s+1},\lambda
_b^{AB^{\prime }}=\lambda _b^{AB}+\frac \Delta {t-j};$ $\lambda
_a^{BC^{\prime }}=\lambda _a^{BC}-\frac \Delta {k-u+1},\lambda
_b^{BC^{\prime }}=\lambda _b^{BC}-\frac \Delta {v-k}$ and the other elements
are the same as before. Since $\Delta $ is sufficient small, we can expand
the function $\ln (\lambda +\frac \Delta K)=\ln \lambda +\frac \Delta {%
K\lambda }$ in the first order$.$ Using this formula, we can calculate the
difference of the function $G$ between these two vectors. 
\begin{eqnarray}
G^{^{\prime }}-G &=&-(Lk-Ni)\lambda _a\ln \lambda _a-(Ll-Nj)\lambda _b\ln
\lambda _b+(Lk-Ni)\lambda _a^{^{\prime }}\ln \lambda _a^{^{\prime
}}+(Ll-Nj)\lambda _b^{^{\prime }}\ln \lambda _b^{^{\prime }}  \nonumber \\
&&+(j-s+1)\lambda _a^{AB}\ln \lambda _a^{AB}+(t-j)\lambda _b^{AB}\ln \lambda
_b^{AB}-(j-s+1)\lambda _a^{AB^{\prime }}\ln \lambda _a^{AB^{\prime }} 
\nonumber \\
&&-(t-j)\lambda _b^{AB^{\prime }}\ln \lambda _b^{AB^{\prime
}}+(k-u+1)\lambda _a^{BC}\ln \lambda _a^{BC}+(v-k)\lambda _b^{BC}\ln \lambda
_b^{BC}  \nonumber \\
&&-(k-u+1)\lambda _a^{BC^{\prime }}\ln \lambda _a^{BC^{\prime
}}-(v-k)\lambda _b^{BC^{\prime }}\ln \lambda _b^{BC^{\prime }}  \nonumber \\
&=&\Delta \ln \frac{\lambda _b\lambda _a^{AB}\lambda _a^{BC}}{\lambda
_a\lambda _b^{AB}\lambda _b^{BC}}  \eqnum{19}
\end{eqnarray}

Since $\lambda _a=\lambda _b$ and $\lambda _a^{AB}<\lambda _b^{AB},\lambda
_a^{BC}<\lambda _b^{BC},$ there will be $G^{^{\prime }}-G<0.$ This is
inconsistent with the suppose that the function $G$ get the minimum. QED

{\bf Fact 5:} When the function $G$ get the minimum, there are at most one $%
\alpha _i^1$ and one $\alpha _j^2$ are nonzero and the elements $\lambda
_k=0 $ ($k\leq Ni$ or $k\leq Lj$), $\lambda _k^{AB}=0$ $\left( k\leq
i\right) ,\lambda _k^{BC}=0$ $\left( k\leq j\right) .$

The proof of this fact is similar to the proof of the second part of the
fact 4. If there is another nonzero parameter, we can take some small value
from the left of this parameter to the right of it to make the value of the
function $G$ smaller, which is inconsistent with the minimal suppose of the
function $G$. This fact means that all of the nonzero elements of the vector 
$\lambda $ are equal to each other. If there is no nonzero parameter act on
the vector $\lambda $, that is, all of the parameters $\alpha _i^1$ and $%
\beta _j^1$ are zeroes, then all of the elements of the vector $\lambda $
are $\frac 1{LMN},$ all of the elements of the vector $\lambda ^{AB}$ are $%
\frac 1{LM},$ all of the elements of the vector $\lambda ^{BC}$ are $\frac 1{%
MN},$ all of the elements of the vector $\lambda ^B$ are $\frac 1M.$ Now the
value of the function $G$ is zero. If there is only one parameter (such as $%
\alpha _i^1$) is nonzero, we have the following fact.

{\bf Fact 6:} When the function $G$ get the minimum and there is only one
parameter $\alpha _i^1$ $(\beta _j^1)$ is nonzero, then all of the nonzero
elements of the vector $\lambda ^{BC},\lambda ^{AB}$ and $\lambda ^B$ are
equal to each other.

{\bf Proof: }Without loss of generality, we suppose the nonzero parameter is 
$\alpha _i^1.$ The nonzero elements of the vector $\lambda ^{AB}$ is equal
to each other. We can get this result by only using the inequality between
the elements of the vector $\lambda ^{AB}$ and $\lambda .$ We focus on the
other part of the fact. Since the nonzero elements of the vector $\lambda
^{AB}$ are the same, all of the parameters $\alpha _i^2$ are zero. Now we
only consider the parameters $\beta _j^2.$ Suppose the nonzero parameters $%
\beta _{_{i_1}}^2,\beta _{i_2}^2,\cdots ,\beta _{i_k}^2$ are set as the
figure II. From the constraints of the elements of the vector $\lambda ^B$
and $\lambda ^{BC}$ in equations (9) 
\begin{eqnarray}
-\ln \lambda _i^{BC}-1-\sum_{j=[\frac{i-1}N]+1}^{M-1}\beta
_j^2+v_{i-1}-v_i+a_2 &=&0\text{ }(i=1,2,\cdots ,MN),  \eqnum{20.1} \\
\ln \lambda _i^B+1+\sum_{j=i}^{M-1}\beta _i^2+w_{i-1}-w_i+a_3 &=&0\text{ }%
(i=1,2,\cdots ,M).  \eqnum{20.2}
\end{eqnarray}
Then we find the elements of these vectors can be divided into several
groups, in each group the elements are equal to each other, that is, 
\begin{eqnarray}
\lambda _{Ni_1+p_1+1}^{BC} &=&\lambda _{Ni_1+p_1+2}^{BC}=\cdots =\lambda
_{MN}^{BC}\equiv \zeta _0^{BC},  \nonumber \\
\lambda _{Ni_1-q_1}^{BC} &=&\lambda _{Ni_1-q_1+1}^{BC}=\cdots =\lambda
_{Ni_1+p_1}^{BC}\equiv \zeta _{01}^{BC}  \nonumber \\
\lambda _{Ni_2+p_2+1}^{BC} &=&\lambda _{Ni_2+p_2+2}^{BC}=\cdots =\lambda
_{Ni_1-q_1-1}^{BC}\equiv \zeta _1^{BC},  \eqnum{21} \\
&&\vdots  \nonumber \\
\lambda _1^{BC} &=&\lambda _2^{BC}=\cdots =\lambda _{Ni_k}^{BC}\equiv \zeta
_k^{BC},  \nonumber
\end{eqnarray}
and 
\begin{eqnarray}
\lambda _{i_1+1}^B &=&\lambda _{i_1+2}^B=\cdots =\lambda _M^B\equiv \zeta
_0^B,  \nonumber \\
\lambda _{i_2+1}^B &=&\lambda _{i_2+2}^B=\cdots =\lambda _{i_1}^B\equiv
\zeta _1^B,  \eqnum{22} \\
&&\vdots  \nonumber \\
\lambda _1^B &=&\lambda _2^B=\cdots =\lambda _{i_k}^B\equiv \zeta _k^B. 
\nonumber
\end{eqnarray}

We must note that all of the parameters $w_i$ which act on the vector $%
\lambda ^B$are zeroes. At first, if all of the indexes $i_j$ satisfy $%
w_{i_j}=0,$ using the fact 1, all of the parameters are zero. The second, if
there are some indexes (such as $i_j)$ make the parameter $w_{i_j}$ to be
nonzero. Because the elements in the same section are equal to each other
for the fact 3, we get $\zeta _{j-i}^B=\zeta _j^B.$ Because of $\zeta
_{j-1}^B\geq N\zeta _{j-1,j}^{BC}$ and $\zeta _j^B\leq N\zeta _{j-1,j}^{BC},$
then $\zeta _j^B=N\zeta _{j-1,j}^{BC}.$ So if we let $l=[\frac{q_j}N]$ and $%
m=[\frac{p_j}N],$ we can get the inequality $\zeta _{j-1}^B\geq (p_j-m)\zeta
_{j-1,j}^{BC}+(N-p_j+m)\zeta _{j-1}^{BC}$ and $\zeta _j^B\leq (q_j-l)\zeta
_{j-1,j}^{BC}+(N-q_j+l)\zeta _j^{BC}.$ Since $\zeta _{j-1}^{BC}\geq \zeta
_{j-1,j}^{BC}\geq \zeta _j^{BC},$ we can get that $\zeta _{j-1}^{BC}=\zeta
_{j-1,j}^{BC}=\zeta _j^{BC}.$ Now we can get the conclusion by using the
fact 1, that all of the parameters $v_k$ $(i_j-q_{i_j}\leq k\leq
i_j-p_{i_j}) $ and $\beta _{i_j}^2$ are zeros. So the second situation can
be reduced to the first situation. So The constraints of the vectors $%
\lambda ^B$ and $\lambda ^{AB}$ are reduced to 
\begin{eqnarray}
-\ln \lambda _i^{BC}-1-\sum_{j=[\frac{i-1}N]+1}^{M-1}\beta
_j^2+v_{i-1}-v_i+a_2 &=&0\text{ }(i=1,2,\cdots ,MN),  \eqnum{23.1} \\
\ln \lambda _i^B+1+\sum_{j=i}^{M-1}\beta _i^2+a_3 &=&0\text{ }(i=1,2,\cdots
,M).  \eqnum{23.2}
\end{eqnarray}

If let $\frac{\zeta _i^{BC}}{\zeta _0^{BC}}=\chi _i$,$\frac{\zeta
_{i-1,i}^{BC}}{\zeta _0^{BC}}=\chi _{i-1,i}$ and $\frac{\zeta _i^B}{\zeta
_0^B}=\eta _i$ $(i=0,2,\cdots ,k)$, we can get $\chi _i=\eta _i$ and $\chi
_{i-1,i}=\chi _{i-1}^{\omega _i}\chi _i^{\sigma _i}$ where $\omega _i+\sigma
_i=1$ and $\omega _i=\frac{p_i}{p_i+q_i}$, $\sigma _i=\frac{q_i}{p_i+q_i}.$
From these definition, we find that all of the parameters $\chi _i$ and $%
\chi _{i-1,i}$ are in the section $[0,1]$. Since we have the conditions $%
\sum_{l=Ni_j+1}^{Ni_{j+1}}\lambda _l^{BC}=\sum_{l=i_j+1}^{i_{j+1}}\lambda
_l^B$, then we can get the equations $\frac{\sum_{l=Ni_j+1}^{Ni_{j+1}}%
\lambda _l^{BC}}{\sum_{l=Ni_1+1}^{MN}\lambda _l^{BC}}=\frac{%
\sum_{l=i_j+1}^{i_{j+1}}\lambda _l^B}{\sum_{l=i_1+1}^M\lambda _l^B}.$ That
is 
\[
\frac{\lbrack N(i_j-i_{j+1})-p_{j+1}-q_j]\chi _j+p_{j+1}\chi _{j+1}^{\sigma
_{j+1}}\chi _j^{\omega _{j+1}}+q_j\chi _j^{\sigma _j}\chi _{j-1}^{\omega _j}%
}{MN-Ni_1-p_1+p_1\chi _1^{\sigma _1}}=\frac{(i_j-i_{j+1})\chi _j}{M-i_1}. 
\]
So we can get the equations 
\begin{equation}
(i_j-i_{j+1})p_1(1-\chi _1^{\sigma _1})=(M-i_1)[p_{j+1}(1-(\frac{\chi _{j+1}%
}{\chi _j})^{\sigma _{j+1}})+q_j(1-(\frac{\chi _{j-1}}{\chi _j})^{\omega
_j})].  \eqnum{24}
\end{equation}

If there is a parameter $w_{Ni_{m+1}}=0,$ then the $m$th equations in
equations (24) has no item which is including $p_{m+1}$. According to the
number of the parameters which make $w_{Ni_{l+1}}=0$ $(1\leq l\leq k)$, we
can divide the elements of these vectors into some sections, the last
equation of this section has no item which contains $p$. We can only point
out that the parameter $v_{Ni_n}$ must be zero where the parameter $n$
satisfy the condition $\eta _n=0$ and $\eta _{n-1}>0$. Or the condition 4
will not be satisfied. So we always can sum up all of the equations in the
same section to get 
\begin{equation}
(i_1-i_{s+1})p_1(1-\chi _1^{\sigma _1})=(M-i_1)[q_1(1-(\frac 1{\chi _1}%
)^{\omega _1})+\sum_{l=2}^s(p_l+q_l)(1-\omega _l(\frac{\chi _l}{\chi _{l-1}}%
)^{\sigma _l}-\sigma _l(\frac{\chi _{l-1}}{\chi _l})^{\omega _l})] 
\eqnum{25}
\end{equation}
where the parameter $s$ means that the parameter $v_{i_{s+1}}=0.$ We first
focus on the lhs. of the equation (25), and obviously, it is non-negative.
Then we consider the rhs. of this equation, there is a function $%
f(x)=xa^{1-x}+(1-x)a^{-x}.$ The value of this function is not more than 1.
Then the rhs. is non-positive. To make the equation to be hold, the two
sides of the equation must be zero. That is $\chi _1=1$ and $\sigma
_i(\omega _i)=0$ or $a=1.$ For each section, we can get the same conditions
which imply that all of the nonzero elements of the vector $\lambda ^B$ and $%
\lambda ^{BC}$ are equal to each other. QED

For the case there are two nonzero parameters $\alpha _i^1$ and $\beta _j^1,$
using the similar method before and notice the condition 4, we can get the
same result that all of the nonzero elements are equal to each other.

For the facts proved before, we can get the conclusion that when the
function $G$ get the minimum, all of the nonzero elements of the vectors $%
\lambda ^{AB},\lambda ^{BC},\lambda ^B$ and $\lambda $ are equal to each
other$.$ Using the condition 4, we can calculate that the minimum of the
function $G$ is not less than zero. This is the end of the proof of the
theorem. QED

Since the Lemma 1 and Lemma 2, the theorem imply that the SSA is hold. This
method can be used to prove some other entropy properties between the
partial density matrix and the multipartite density matrix, Such as the
inequality $S(\rho _{AB})\leq S(\rho _A)+S(\rho _B).$

\section{Conclusion}

In this paper we give a new elementary proof of the SSA which is an
important property of the entropy for classical information and quantum
information. The proof is dependent on the analysis of the minimal value of
a function under some conditions. This proof also show that the conditions
in our previous work{\it \ }\cite{han} are stronger than the SSA.

\section{Acknowledge}

This work was funded by the National Fundamental Research Program
(2001CB309300), National Natural Science Foundation of China, the Innovation
Funds from Chinese Academy of Sciences, and also founded by the outstanding
Ph. D thesis award and the CAS's talented scientist award rewarded to
Lu-Ming Duan.

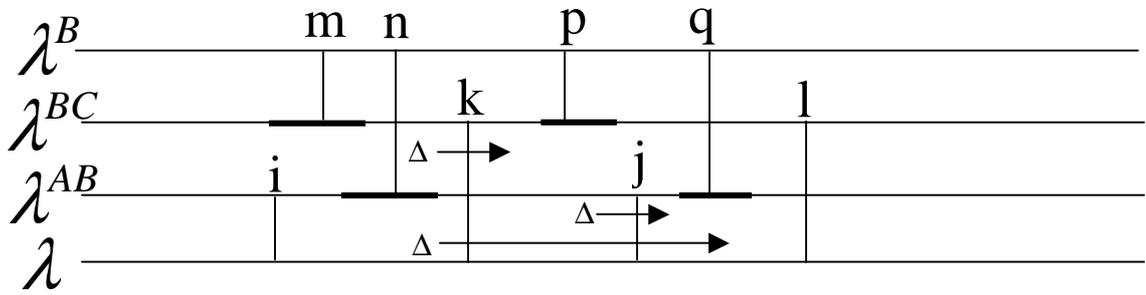
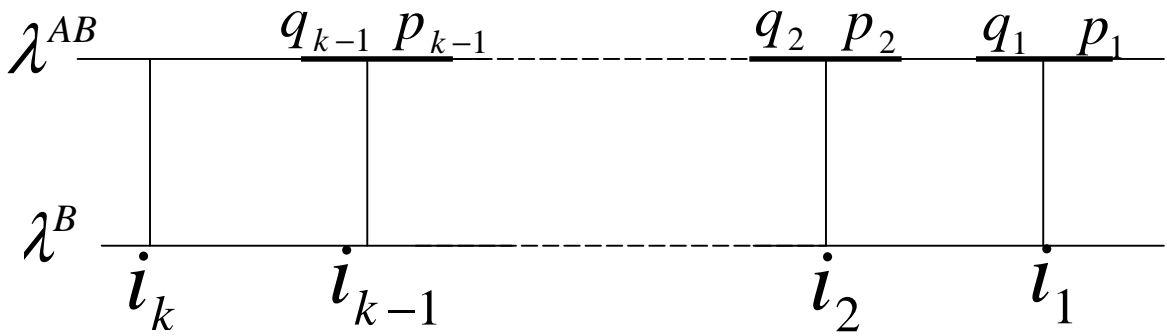
\end{document}